\documentclass[
aps,%
12pt,%
final,%
notitlepage,%
oneside,%
onecolumn,%
nobibnotes,%
nofootinbib,%
superscriptaddress,%
noshowpacs,%
]%
{revtex4}
\usepackage{graphicx}

\begin{document}

\title{Optical Counterparts of the X-ray Sources Hercules X-1 and
Cygnus X-2: Genuine and Fake}
\author{\firstname{V.~P.}~\surname{Goranskij}}
\affiliation{Sternberg State Astronomical Institute, Moscow State
University, Moscow, 119992 Russia} \email{goray@sai.msu.ru}

\begin{abstract}
This paper is a refutation of two researches by A.~N.~Sazonov on
HZ~Her and V1341~Cyg published in the Astronomy Reports (Vol.~55,
p.p.142 and 230) in 2011. We analyzed his photometry along with other
data collected in the literature and Internet.
The observations of Sazonov are fake and are not related to real stars.
Conclusions contained in his papers do not constitute any scientific value.
We describe the properties of these two objects, derived from real
observations. The following papers based on the same fake observations
and published by him in astro-ph arXiv 0904.0168, 0907.3822, 0912.0706,
1011.3980, and 1102.0379 are also inauthentic.
\end{abstract}

\maketitle

\section{INTRODUCTION}

Her X-1 is an eclipsing X-ray  system, consisting of a neutron
star with the mass of about 1.4 M$_{\odot}$ and a star of ~A7 spectral
type with a mass of 2.2 M$_{\odot}$, which fills its Roche lobe and is
an accretion donor~\cite{Jur01}. In the optical range the star
is known as HZ~Her. The orbital period of the system is 1.7 days. A wave with
this period dominates in the optical light curve. This
large-amplitude wave (1$^m$.7 in the $B$-band) is formed due to the
reprocessing of X-rays coming from the neutron star on the surface
of the donor, facing the neutron star (reflection effect). The
neutron star is an X-ray pulsar with a period of 1.24~s.

Cyg~X-2 is also an X-ray system, which includes a neutron star
with a mass of 1.7 $\pm$ 0.2~$M_\odot$ and an optical component -- an
accretion donor of the spectral class A5 - F2~III, whose mass is
estimated to be 0.6 $M_\odot$~\cite{Cas02}. The optical counterpart of this
system is known as V1341~Cyg. The inclination of the orbit to the
line of sight is 62.5 $\pm 4^\circ$, and the orbital period amounts to
9.84450~days. Since the radial velocity  of the system $\gamma$ =
--210 km/s, the star is the object of the spherical component of Galaxy
population, and based on its photometric properties its
optical component is likely to be a blue straggler~\cite{Gor03}.

In 2011, the Astronomy Report (Russian Astronomicheskii Zhurnal)
published two papers by A.~N.~Sazonov describing the multi-color
$WBVR$ photoelectric
observations of the optical counterparts Her~X-1~\cite{Saz04} and
Cyg~X-2~\cite{Saz05}. The first glance at Figs.~1 and 2 in~\cite{Saz04}
gives doubts about the authenticity of the observations of HZ~Her, since the
light curves in different filters are lacking the primary eclipse
with a "flat bottom." It is known that in the primary eclipse of
HZ Her the occultation of the accretion disk occurs.
Usually in certain phases of the eclipses, we observe
characteristic brightness variations, that occur at the
ingress of the precessing disk behind the companion star, and egress
from behind it. These brightness variations occur at different rates and
depend on the "on" and "off" state of the X-ray source there~\cite{Liu06}.
The study of ~Sazonov~\cite{Saz04} did not reveal such brightness
variations. Similar doubts arise on a closer examination of the
V1341~Cyg light curves in Fig.~1 of~\cite{Saz05}, where all the filters
clearly show the eclipse-like minima: a deep and narrow minimum
around the orbital period phase 0.0, and a shallow minimum around the
phase~0.5. So far, this system with a small orbit inclination has
only revealed the ellipsoidal brightness variations of the secondary
component, which were superimposed by
the chaotic variability. The amplitude of the chaotic variability
increased from the visible to the ultraviolet range~\cite{Gor03}.

The eclipse in the system of HZ~Her, the presence of a neutron star
and an accretion disk surrounding the neutron star and precessing
with a period of 35 days are firmly established facts, with
the participation of independent observers working in different
ranges of the electromagnetic spectrum. The star has also revealed
abnormal states. It is known that the system was inactive for
seven years between 1934 and 1940, when it lacked the reflection
effect and the  ellipsoidal effect of its optical component~\cite{Wen07}.
In this event, the star remained eclipsing, and its light curve
revealed two brightness minima with the depth of about $0^m.5$. HZ~Her
was reported to have cases of abnormally low states in X-rays
(ALS), when the flux was very weak and did not reach its normal
high level in the "on" state.
Abnormally low states lasted for several 35-day
cycles~\cite{Jur01}. However, the optical light curves in the ALS were not
different from those observed in the normal "on" state. Abnormal
states with brightness levels lower than in the quiescent state,
which is characterized by an ellipsoidal light curve were also
observed in visible light in V1341~Cyg~\cite{Gor03}. The possibility of
occurrence of abnormal states has to be taken into account when
considering the observations of A.N.~Sazonov.

To test Sazonov's observations, we compared his tabular data, put
out in the Internet and accessed from the references in~\cite{Saz04} and
\cite{Saz05} with the observations of other authors and with the
observations of the SuperWASP-N robotic telescope~\cite{Col08}, mounted on
the La Palma island (Roque de los Muchachos Observatory, Canary
Islands, Spain). The data on HZ~Her used for comparison, is a
collection of photoelectric multicolor observations in the $U(W)BVR$
bands in the time interval JD 2441511 -- 2451046 (1972 -- 1998).
This database was analyzed earlier by N.~I.~Shakura et al.~\cite{Sha09}.
It is supplemented by the unpublished observations of V.~M.~Lyutyi and
I.~B.~Voloshina, it contains 5936 observations in the $B$ band and a
smaller number of observations in the $U(W)$, $V$ and $R$ bands. The
database~\cite{Sha09} was amended: the magnitudes were corrected to bring the
discordant observations in the same photometric system, and some
clearly erroneous observational nights were eliminated. The
database does not contain Sazonov's earlier published observations
\cite{Saz10}. Sazonov's tables~\cite{Saz04} contain 1808 observations
covering the period of JD 2446588 -- 2448173 (1986 -- 1990) and overlapping in
time with the database in~\cite{Sha09}. This time interval~\cite{Sha09}
contains 977 observations, which were made by independent groups of
observers: Lyutyi and Voloshina~\cite{Lyu11} and Kilyachkov et al.~\cite{Kil12}.
On November 2, 2011 the table of observations of HZ~Her was changed
by Sazonov, when 133 observations were deleted from it and 42
observations were added. The new table is also analyzed in the
present paper. The SuperWASP-N observations were  obtained in the
period from May 2004 to August 2007 (JD 2453129 -- 2454681), and
their number amounts to 41942. These observations have previously
been analyzed in~\cite{Jur01}. The system of these observations is reduced
to $V$, the time is measured in seconds from JD = 2453005.5, and
radiation fluxes are given in the  Micro-Vega units,
related to magnitudes via the formula $mag = 15 - 2.5\ log (Flux)$.
The accuracy of these observations differs a lot, and therefore
from the whole data set we have selected 32143 observations, the
accuracy estimate of which does not exceed $0.08^m$. The nights
with conflicting data were removed. It was easy to determine the
cause of the controversy: observations were made with two cameras
installed at the same mount, the object is located in different
parts of the cameras' field of view, simultaneous observations are
not tied to the comparison stars, and are hence systematically
different. The time points of the SuperWASP tables are converted
into Julian dates, and fluxes -- in magnitudes.

To establish the reliability of Sazonov's data on the object
V1341~Cyg~\cite{Saz05}, we used the photoelectric $UBV$ observations from his
earlier work~\cite{Saz13}, photographic observations of M.~M.~Basco et
al.~\cite{Bas14} in the $B$ band, photoelectric $UBV$ observations of
V.~M.~Lyutyi from~\cite{Gor03,Bas15}, and CCD observations of the
author in the $UBV$ filters, performed with the 1-m Zeiss-1000 reflector
of the SAO RAS, the Maksutov 50-cm meniscus telescope  and the 60-cm
reflector of the Sternberg Astronomical Institute Crimean
Station. The photographic observations~\cite{Bas14} were obtained with the
Maksutov 50-cm telescope and feature the microphotometric
measurements with an iris photometer. The full-time span of these
observations is limited by the Julian dates JD 2442247 -- 2455826
(1974 -- 2011). The observations from~\cite{Gor03,Bas14,Bas15} and CCD
photometry have been reduced to a single photometric system of the SAO RAS.
A collection of observations of the author is available for viewing
with a Java-compatible browser on the Internet at \begin{verbatim}
http://jet.sao.ru/~goray/v1341cyg.htm,\end{verbatim} the table of observations
is contained in the file \begin{verbatim}
http://jet.sao.ru/~goray/v1341cyg.ne2.\end{verbatim}
The first column of the given
table sets time points in the JD hel.-- 2400000.0 format, the four
subsequent columns contain observations in the $V, B, U,$ and $R$
order, and the last column contains the mark of two letters or
numbers indicating the observational data source. At the time of
analysis the file contained 1749 observations in the $B$ band, 743
sets in the $V$ band and 589 - in the $U$ band. Sazonov's table
from~\cite{Saz05} contains 296 night-averaged $WBVR$ values in the time
interval of JD 2446616 -- 2448521. The number of observations of
V1341~Cyg by the robotic telescope SuperWASP-N is 6444. The date
of the countdown and intensity scale is the same as that for
HZ~Her. All the observations of SuperWASP-N were converted to
Julian dates and magnitudes. The sample of observations was
compiled based on the same features, as for HZ~Her. The selected
observations fall in the time interval from June 2004 to November
2007 (JD 2454279 -- 2454419, 88 nights), and their number is 4053.

We have also analyzed the observations of these two X-ray sources
made by the RXTE orbiting observatory with the ASM  camera in the
X-ray range of about $2 - 15$~keV, available at the website
\begin{verbatim}http://xte.mit.edu/ASM_lc.html.\end{verbatim}

\section{Methods of anslysis}

To identify and study the periodic components in the light curves,
we used the methods of Lafler-Kinman~\cite{Laf16} (phase dispersion
minimization, PDM) and Deeming~\cite{Dee17} (Fourier transform of discrete
time series). We applied the EFFECT code, implementing these
methods in the OS Windows environment and written by the author of
this article. The software and user guide are available at \begin{verbatim}
http://vgoray.front.ru/software.\end{verbatim} The first method~\cite{Laf16}
is the most convenient to refine the orbital period of the system HZ~Her, in
which the light curve is dominated by the reflection effect. The second
method~\cite{Dee17} is applicable for the identification of the
ellipsoidal effect wave from the chaotic light variability of
V1341~Cyg, which has a relatively small amplitude. The software
allows to calculate the average light curves of the periodic
components, smooth them by hand and subtract by the phase from the
light curves (prewhitening procedure).

As known, the optical photometry of HZ~Her shows a wave with an
orbital period of 1.70017 day~\cite{Kur18,Tan19}. Residuals after subtracting
the wave reveal the period of 1.62 days. This is the beat period
between the precession period and the orbital period~\cite{Ger20}. The
period of precession of the accretion disk is not detected in the
residuals neither by the methods of dispersion minimization, nor by
the Fourier transform, although the shape of the orbital light
curve varies with this period. The lack of precession frequency in
the residuals is due to the fact that in different orbital phases,
precession brightness variations are either not visible (in the
eclipse) or visible, but are caused by different reasons
(precession of the disc itself, or the drift of the shadow from
the plane of the disc over the stellar surface facing it). For
these reasons, brightness variations with the precession period
are in the phase opposition in different orbital phases. The author
(V.~P.~Goranskij) proposed a computer-based method for finding
periodicity in the light curve shape variations~\cite{Gor21,Gor22} which
makes it possible to identify and refine the periods of
precession, or the light curve amplitude modulations (e.g., for
the RR Lyrae type variable stars with the Blazhko effect). In the
computer search of the periodicity, we usually test the trial
periods in a fairly wide range in rather small steps in frequency,
not to miss the real period value. Searching for periodicity via
the Goranskij's method, the computer memory builds the light curves
with the known primary period (of the orbit or pulsations),
separated by the phase intervals of the secondary trial period
(precession or amplitude modulation). As the dispersion parameter,
similar to the method~\cite{Laf16}, we used the sum of squared deviations
of each light curve point, subsequent in phase of the main period
from the previous one. The parameter, on which we based our search,
is the ratio of dispersion of the total light curve to the
dispersion of  light curves, divided by the phase of the
secondary trial period. When the value of the trial period
approached to the actual value, dispersion in the divided light
curves sharply reduced, resulting in the parameter increase, and a
peak appearing in the periodogram. The method allows to refine
two periods of the system with the precessing disk, both orbital
and precession~\cite{Gor22}. This method was used to determine whether
there exists a precession period in different sets of
observations of HZ~Her.

\section{Results of Analysis}

\subsection{HZ~Her}

An analysis of four series of observations of HZ~Her has revealed
the periods of the orbit P$_{orb}$, of the beats P$_{beat}$,
and precession P$_{prec}$ (Table~\ref{tab1}).
The parentheses give the errors of period
determination in units of the last digit, the abbreviation LK, D,
G imply the following methods of analysis: Lafler-Kinman, Deeming
and Goranskij. To study the period variations we used the O--C
method. The line "JD 24..." shows the time range of the
observational set, and the "N" line gives the number of
observations. An asterisk indicates the periods determined from a
part of observations, the sample was compiled based on the phase of
the orbital period in the ranges of $\varphi = \pm (0.08 - 0.14)$.

The orbital period is most precisely determined from the Doppler
shift of the HZ~Her pulsar pulses in the X-ray range~\cite{Sti23}. The
variation of the orbital period occurred between the values  of
1$^d$.700167790 (10) in the time range of JD 2441300 -- 2445120 and
1$^d$.700167504 (7) in the range of  JD 2445120 -- 2451000. The period
variation may have been smooth in this time range. Such tiny
variations in the orbital period can not be registered by the
photometric method. Orbital periods, determined from all the four
rows, do not contradict with the data of~\cite{Sti23} within the error.
The light curves with the orbital period are shown in Fig.~1
from the data collection by N.~I.~Shakura and colleagues~\cite{Sha09}, b --
according to SuperWASP~\cite{Jur01}, c -- according to A.~Sazonov~\cite{Saz04},
d -- according to the data of the collection of observations in~\cite{Sha09},
only related to the time period, which includes Sazonov's
data~\cite{Saz04}. At the bottom (e) you can see the orbital light curve
according to the RXTE orbital observatory in the 2-15 keV X-ray
range.

The optical light curves of HZ~Her (Fig.~1~a,~b,~d) reveal a
decrease in the dispersion of observations near the orbital phase
$\varphi$ = 0.0 -- eclipses of the accretion disc. Minimum variability
is observed in the largest phase of the eclipse, when the light of
the back side of the companion  (A7-type star) dominates. The
variability is reduced because in this phase the highly variable
brightness of the precessing disk and the hot spot on the side of
the companion facing it is not visible. The data by Sazonov~\cite{Saz04}
(Fig.~1~c) show the equal dispersion during all phases. With the
negligible dispersion of individual rows of monitoring in~\cite{Saz04},
reflecting the high accuracy of his data, the scatter of points in
the zero phase remains maximum. This means that the data in~\cite{Saz04}
lacks any traces of the eclipse of the precessing accretion disc.
Such a behavior of the star in the same period of time is not
supported by the observations from the collection by Shakura et
al.~\cite{Sha09} acquired at the same time (Fig.~1~g).

The X-ray range (Fig.~1~e) near the zero phase reveals a total
eclipse of the source lasting  $0^d.24$. Phases of the beginning and end
of the eclipse are $\varphi  = \pm 0.0706$  relative to the middle of the
eclipse. In this diagram, we do not observe the zero flux in the
phase of the eclipse only because of the insufficient measurement
accuracy of the ASM instrument on the RXTE orbiting observatory, where
the measurement accuracy near the zero level amounts to 2.5 cps.
At a low or zero flux, which is observed during the eclipse and in
other orbital phases in the low "off" state, the counts  prove to
be positive or negative (below zero) with equal probability. When
the eclipse is viewed in the X-ray range, the occultation of the
pulsar appears and pulsating radiation disappears. Before the
eclipse, the phases $\varphi = 0.80 - 0.90$ of the X-ray orbital curve
reveal a large depression because of the numerous short-term X-ray
fall-offs (e.g., described in~\cite{Ign24}), which are explained by the
overlap of the X-ray emission by the thickening at the edge of the
accretion disk (a blob), emerging in the region of the
collision with the accretion stream~\cite{Cro25}. Fast recessions and even
flickering of the X-ray source in some phases of formation of this
blob may be associated with the separation of the blob into
sprays, which move on Keplerian orbits above the disc
plane~\cite{Boc26}. Igna and Leahi~\cite{Ign24} presented a survey of
alternative explanations and patterns of this phenomenon, which is, however,
not observed in the optics.

The secondary eclipse also looks unusual in Sazonov's data~\cite{Saz04},
during which there occurs a transit of the precessing accretion disk in
front of the companion against the hot spot on the companion's
surface (reflection effect). Observations~\cite{Jur01} and~\cite{Sha09}
indicate that the secondary eclipse is not always visible or its amplitude
is small, although it can be deep. The largest depth of the eclipse
is observed in the high "on" state of the 35-day cycle. At the same
time when the secondary eclipse has the greatest depth, the
amplitude of the reflection effect wave decreases. These features
of the light curves of HZ~Her have been well demonstrated in
Fig.~5 by Gerend and Boynton~\cite{Ger20}. In Sazonov's data~\cite{Saz04}, deep
secondary eclipses occur in all the phases of the precession
cycle if these phases are counted from the elements determined
from the observations~\cite{Jur01} and~\cite{Sha09}. It is easy to explain the
appearance of deep secondary eclipses in the "on" state. In these
phases the accretion disk is open to the observer on Earth, and
when passing in front of the secondary component, the surface of
which has a bright hot spot, it covers the largest area of this
spot. Moreover, the shadow of the plane of the disk is visible
separately therefore the area of shadow additionally reduces the
spot brightness. In the "off" state, the disk is visible
"edge-on", and in the orbital motion, as viewed by an external
observer, it moves along its shadow, so the loss of brightness of
the hot spot for this observer is minimum, the eclipses are almost
invisible, and the amplitude of the reflection effect gets maximum.
If the secondary eclipse was in fact always deep, as stated in
Sazonov~\cite{Saz04}, the accretion disk would be always open and only towards
the observer, which is a nonsense and contradicts the X-ray data,
clearly demonstrating the high and low states.

Fig.~2 shows the fragments of the orbital light curve of HZ~Her,
built in the orbital phases near the primary eclipse. During the main
eclipse, the precessing accretion disk, which is a highly variable
object, gets covered. The observations~\cite{Jur01} and~\cite{Sha09}
(Fig.~2~a,~b,~d) near the zero phase show an area of the "flat bottom"
with minimum variability, when the disk is completely covered.
A careful study of
these observations in the "flat bottom" area in different eclipses
sometimes reveals  small-amplitude variations, such as jumps or
trends of the light curves, but these variations are small and are
in fact the radiation of the circumstellar gas, not obscured in
the eclipse.

When the precessing accretion disk comes into eclipse, the
brightness decay rate varies depending on the opening angle of the
disk towards the observer, resulting in a gradual reduction of the
dispersion of observations. When the eclipse is ending, the
dispersion of observations increases for the same reason. In Sazonov's
data~\cite{Saz04}, the brightness variation rates are about the same, the
scattering of certain light curves does not change even near the zero
phase, hence, the eclipse is missing in these phases (at that,
however, a deep secondary eclipse is present). This is nonsense,
from which it follows that the data~\cite{Saz04} is spurious. The main
eclipses were always observed in this system, both in the X-rays
and optics, and even when the reflection effect was off. The
presence of the primary eclipses is confirmed both in the
sample of data~\cite{Sha09} in the same time range, which includes the data
of~\cite{Saz04}.

When the neutron star, which is the X-ray source, gets covered,
the hot spot of the reflection effect becomes invisible for the
terrestrial observer. In the narrow range of orbital phases ~0.03
after the neutron star is covered by the limb of the secondary
component, and in the same phase interval prior to the opening of
the neutron star, the accretion disk is partly visible, and we can
observe its precession from the opposite sides. Prior to the first
contact of the neutron star and after its last contact, a large
part of the accretion disk or the entire disc is visible, but the
contribution of the hot spot proves to be significant (which looks
like a narrow crescent moon in these phases). Near the contacts,
while the contribution of the disk is relatively small, the effect
of the hot spot can be compensated in order to extract the light
curve which is due only to the precession of the disk. To do this,
we subtracted the average orbital light curve by the orbital
phases, and  selected in the excess file the observations from
the orbital phase ranges $0.86 - 0.92$ and $0.08 - 0.14$. From
this sample, from different series of observations, we can
determine and refine the precession period by the Lafler--Kinman
method from the dispersion minimum. For the sample from the
collection of Shakura et al.~\cite{Sha09}, it becomes equal to
$34^d.89 \pm0^d.01$ (Fig.~3~a), and
from the sample of SuperWASP data~\cite{Jur01} it amounts
to $34^d.75 \pm0^d.05$ (Fig.~3~b). In the sample of these orbital
phases, we can observe the precession of the accretion disk "in
its purest form". The light curve, constructed in accordance with
the precession phase, shows two peaks: a high peak - opening the
disc "from above", and a low peak - opening the disc "from
below". In the corresponding sample of Sazonov's data~\cite{Saz04} this
period is completely lacking (Fig.~3~c).

The stability of the precession period (the accuracy of the
"35-day clock") has been studied in~\cite{Boy27} from the X-ray data.
The O--C curves were determined from the time of switching the
X-ray source "on", from the time of absorption recession occurring
in the  X-ray light curve and  from the optical light curves via
the method proposed by J.~Deeter~\cite{Boy27}. According to Deeter, the
precession phase can be refined from the three-dimensional phase
diagram orbital phase--precession phase--deviation from the mean
orbital light curve. All these methods have shown such variations
of the precession cycle that the O--C deviations vary within 6--7
days, i.e. up to 19\% by the precession phase~(\cite{Boy27}, Fig.~1).
Fig.~3~a,~b show the way the light curve is floating exactly
within the same  precession period phase interval. However, the
graphs show the precession curve with two brightness peaks: with a
maximum around $\varphi =  ~ 0.15$, corresponding to the high "on"
state, and with a maximum around $\varphi = ~0.65$, corresponding to the
low "on" state in the X-rays. These states correspond to the 
opening of the accretion disk from the top and the bottom.

The stability of the precession period of HZ~Her from the RXTE/ASM
X-ray data was studied in~\cite{Lea28}. Precession variations of the X-ray
flux are observed in all phases of the orbital period, except for
the primary eclipses. In the X-ray range the light curve of the
precession period is not disturbed by the influence of the
reflection effect and the movement of the shadow on the
companion's surface, and thus resembles the curve in the visible
range that we have identified in the narrow range of orbital phases
(Fig.~3~a,~b). The precession phase curve has the form of a double
wave with high and low "on" states, and in the "off" states the
flux is near the detection level~(\cite{Lea28}, Fig.~2). The RXTE/ASM
satellite recorded two abnormally low states (ALS), at which the
precession curve with a double wave was absent, and the radiation
flux remained at a low level:  JD~2451280--2451647 (April
1999--March 2000) and JD~2453003--53091 (January -- March 2004).
The X-ray orbiters RXTE/PCA and BeppoSAX have registered in the state
of ALS in 1999--2000 the X-ray emission of the pulsar, reflected by
the secondary component. The intensity of the pulsar varies with
orbital period~\cite{Sti29,Par30}. In the optical range there are 445 NSVS
database  observations for this ALS episode \begin{verbatim}
http://skydot.lanl.gov/nsvs/nsvs.php,\end{verbatim} which confirm that at this
time the reflection effect was present in the visible range, and
its amplitude was $0^m.7$ (CCD without the filter). These
observations suggest that direct X-ray emission in the ALS was
blocked in the direction of the terrestrial observer as a result of
changes in the structure of the disk, while the central X-ray
source had a normal luminosity and irradiated the surface of the
companion. However, no optical or X-ray observations provide
information about whether the precession of the disk persisted
during the ALS. Our O--C analysis of the 35-day periodicity
according to the up-to-date RXTE/ASM data in the range of \mbox{JD
2450087--2455817} shows the precession period variability between
$34.5 - 35.2$ days. This causes phase shifts of the precession curve
with the amplitude of 13 days with respect to the linear elements,
what is demonstrated in Fig.~3~d.

The analysis of periodicity in the residuals after subtracting the
mean light curve of the orbital period (the reflection effect) was
carried out using the Deeming method, and its results are summarized
in the third row of Table~\ref{tab1}. In the data of~\cite{Jur01} and~\cite{Sha09}
the beat period $P_{beat} = 1^d.621$ is present, coupled with the precession
period $P_{prec} = 34^d.9$ and orbital period as follows:\\
          
           $P_{prec}^{-1} = P_{beat}^{-1} - P_{orb}^{-1}$.\\

\noindent
Sazonov's data~\cite{Saz04} lacks the beat period.\\ 

The precession period was found and refined from the optical
observations of~\cite{Jur01} and~\cite{Sha09} by Goranskij's method,
including all phases of the orbital period. The results are
given in Table~\ref{tab1}, and the periodograms are shown in Fig.~4.
These calculations have shown that the period of precession in the data of
Sazonov~\cite{Saz04} is missing absolutely. However, the precession period
is present in~\cite{Jur01} and~\cite{Sha09}, it is also present in the
sample from the collection of~\cite{Sha09} in the same time range to which
pertains the data in~\cite{Saz04}.
The precession period  is detected despite its variability and
small shifts of the  precession light  curves by phase.
This  fact indicates that Sazonov's data are fake, and  precession
brightness variations were not included during their preparation.

Comparing the collection by Shakura~et al.~\cite{Sha09} with Sazonov's data~
\cite{Saz04}, 
we found two nights during which the long monitoring series coincide in
time, and which belong to the phases when the accretion disk leaves
the primary eclipse in the 'on' state. These are the nights JD 2447365
and 2448062. During these nights Sazonov's data~\cite{Saz04}
coincide in time with the observations of Kilyachkov et al.~\cite{Kil12}. 
The light curves in  $B$ and $U$ ($W$) filters are compared in Fig.~5.
The differences between the data of~\cite{Sha09} and~\cite{Saz04} in the $B$
filter reach $0^m.3$, and in the ultraviolet filters they are up to $1^m.0$.
The differences in the ultraviolet
light curves of the $U$ and $W$ filters can not be explained by the
differences in the filter response curves only. The reason of such
large differences is obvious: in the process of counterfeit,
Sazonov's data failed to take into account the precession
variations of brightness. Due to the falsification  
it is not advisable to use the data~\cite{Saz04} for the study of HZ~Her.

\subsection{V1341~Cyg}

The orbital period of V1341~Cyg is known from the spectroscopic
observations. Its latest and the most reliable determinations 
belong to Elebert et al.~\cite{Ele31}, $9^d.84456 \pm0^d.00012$, and Casares et
al.~\cite{Cas02}, $9^d.84450 \pm0^d.00019$. We used the latter value for the
construction of  light curves. The initial point of phase count we used 
was the epoch of the inferior conjunction of the normal class F2  
star T$_0$ = JD 2451387.148 $\pm$0.018 from~\cite{Cas02}.

The light curves, constructed with these elements are shown in Fig.~6. 
All the scales in these diagrams are the same. To the left you can see
a collection of observations compiled by the author (referred to as the
Goranskij's collection).
These observations show a double wave over the orbital period on
the lower, quiet brightness level and a significant chaotic 
variability, which increases from short- to long-wave filters
(from the $V$ filter to the $U$ filter).
The amplitude of the wave is on  the average $0^m.27$ in
the $V$ filter. In~\cite{Gor03}, this wave was explained by the ellipsoidal
effect. New observations made in the SAO and SAI do not contradict
the old photoelectric observations by V.~M.~Lyutyi and old photographic
observations, showing the characteristic behavior of the object.
Active and quiet states are typical of it. 
In the active states  brightness increases, and the points deviate
up from the double wave. 
The right-hand side shows the light curves in the $WBVR$ system, built
according to Sazonov. They have the same shape in all spectral bands, 
reminiscent of the light curves of the eclipsing contact
systems such as W UMa. One can not distinguish quiet and
active states in these data. The amplitude of the orbital variability
is on the average $0^m.6 V$, which is twice the amplitude of the
ellipsoidal effect from the Goranskij's collection data. This amplitude
is too large to be explained by the ellipsoidal effect. 
Attention is drawn to the same ratio of the depth of minima in the 
light curves in different filters. Since the system has an accretion disk,
which makes a significant contribution to the total brightness of the system
and has an ultraviolet excess, the depth of the primary eclipse in
the $W$ band  has to be much larger than in the other filters, what is 
not visible Fig.~6. Sazonov~\cite{Saz05} concluded that the
optical star in the V1341~Cyg system belongs to red giants, rather
than to the "blue stragglers" (despite the fact that the spectrum of
the optical star is clearly visible in the total spectrum of the system
and is defined as the A5-F2 III, and it is not a red giant). The light
curves, based on Sazonov's "multi-color observations", which
show the same ratio of the depths of minima in different
filters, are in an insuperable conflict with his conclusion about the red
giant.  In addition, the orbital periods of the X-ray systems with  
red giants (SyXB) make up hundreds of days, while in V1341~Cyg
the orbital period is only $9^d.84$.

The results of the periodicity analysis  of V1341~Cyg are listed
in Table~\ref{tab2}. According to the
observations from the Goranskij's collection in the ultraviolet, the orbital
period can not be detected by any frequency analysis method.
However, the average light curve in the $U$ band with accepted
elements reveals a wave with the amplitude of $0^m.2$ and a
brightness minimum, which coincides with the moment of inferior conjunction
of the optical component.
This is quite certainly a reflection effect similar to the 
reflection effect, observed in HZ~Her, whose amplitude is much smaller
due to the small inclination of the orbital plane to the line of
sight. The reflection effect usually
has maximum amplitude in the ultraviolet rays. The peaks with a period 
equal to half the orbital period are reliably detected in the $B$ and $V$
filters, applying the Deeming method.
The Lafler--Kinman method detects both the double wave with the orbital
period, and a solitary wave with a half-period. Sazonov's data~\cite{Saz05}
contain a periodic component (a double wave for the
period), which is present in all filters. The differences of the orbital
period according to~\cite{Saz05} and spectroscopic~\cite{Cas02} are
not significant.

In this paper we also analyzed the observational data of the SuperWASP and
RXTE orbital observatories. In the SuperWASP optical data
(column FLUX2) the object mostly varies in the range of values
$14.0 - 14.9$ in the $V$ filter,
although a number of (several tens of) observations fall below, up to the
value of $15^m.6$. These flux decays are mainly concentrated in phases
$-0.3 < \varphi < +0.1$. They are mostly random, though some are clearly
related to the breaks or the end of monitoring. Such profound
light decays are absent both in the old photoelectric observations by Lyutyi
and in the modern CCD observations. These decays are
perhaps associated with passing clouds. According to the SuperWASP, the
double wave with the orbital period of $9^d.82$ and the average full amplitude
of $0^m.13$ is confirmed.

The RXTE/ASM data reveal  a lot of flares and active states, so that the
level of the quiescent state can not be determined. A multitude of
short-term flux dips can be observed, which occur in different phases
of activity, and even near the flare maxima. They are
not associated with the orbital phase. So Cyg~X-2 is not an eclipsing
X-ray source. The frequency analysis does not detect any periodic
variations near the orbital period, even after the Fourier
decomposition and the cleanup of this series from the long-period components.

The differences in the shapes of light curves of
V1341~Cyg, built based on the observations from the early study of
Sazonov~\cite{Saz13} (which are contained in the Goranskij's collection) and
based on the data from the work of Sazonov~\cite{Saz05} are obvious. 
Some of these observations  overlap in time. In Fig.~7~a the 
observations from~\cite{Saz13} are circled on
the background of Goranskij's collection data (points). They are in a 
very good agreement with the photoelectric observations by Lyutyi and
CCD photometry. Fig.~7~b compares the observations of Sazonov
from~\cite{Saz13} with the observations of Sazonov from~\cite{Saz05}
depending on time. The dates of observations from~\cite{Saz13}
and~\cite{Saz05} coincide, which makes it evident that
the data in~\cite{Saz05} have been amended, which resulted in an artificial
enlargement of the amplitude of orbital
variability, and attaching the narrow "eclipses" that have previously
never been observed by anyone, including Sazonov himself.
This concocted part of the observations from~\cite{Saz05}, as
well as the other observations from the same source can not be
considered reliable. The conclusions of paper~\cite{Saz05}, based on unreliable
data can therefore not be considered objective and valuable for science.

The papers of Sazonov, published in arXiv.org~
\cite{Saz32,Saz33,Saz34,Saz35,Saz36} and based on the same data should also
be considered unreliable. The use of these materials for the scientific
research purposes is not recommended. However the reliability of the
earlier observations~\cite{Saz10,Saz13} as well as the observations
of other objects published by him before 1996 can not be doubted.

At the same time, the  SuperWASP data contain a large percentage
of defective and unreliable observations, and they should be used
with caution. Obviously, these are absolute measurements and an
account of the variable atmospheric extinction over large areas of
the sky on the CCD frames can lead to significant systematic
errors. These data can be perhaps improved for individual stars
using the measurements of the nearest neighbor stars as a
standard.

\section{Additional remarks}

Sazonov's observations in~\cite{Saz04} and~\cite{Saz05} have been presented
in the PhD thesis, and were therefore considered by the Dissertation Council
D501.001.86 of Sternberg Institute of the Moscow University
and a specially appointed Commission of this council
(Transactions No.103 on 15 December, 2011).

The Commission found that "the observational data sets~[series] that satisfy
a linear function of time with the accuracy, ensuing from the author's
numerical data could not have been obtained in the stated conditions
and using the tools available to the thesis' author." This applies to
the data on HZ~Her~\cite{Saz04}. In other words, it is recognized that the
accuracy of $0^m.001$ in the color indices for a star having
the brightness of $14^m.8$, is not real. The presence in Sazonov's data
of small time intervals between the observations of different objects, 
eliminating the possibility of telescope repointing was confirmed.
These facts are indicative of the falsification of the observational data.

At the same time, examining Sazonov's data, the Commission judged
from the following provisions: (1) "the discrepancy of the observational
results of the defender of thesis with other observational results
does not allow to consider them obviously incorrect"; 
(2) "the discrepancy of the observational results of the defender
of thesis with the
currently existing models of the studied objects does not preclude
from considering them [the observations] as obviously incorrect."
Based upon these provisions, the facts set out in this paper do not prove
the counterfeit of the original observational data in Sazonov's papers.

Such initial assumptions of the Scientific Council on the thesis
defense are quite puzzling. 
On the cosmic scale human life and even the existence of mankind are
inappreciable. Each event in the astrophysics may be prove to
be unique or never recur again for centuries. The reliability of data
in the astrophysics can be tested and determined only in the independent
observations by different observers with different tools and
instruments. No other possibility exists in the astrophysics, since
it is impossible to carry out the experiment. 
The only exceptional opportunity to look into the past is the
light echoes of large explosions. 
Special means are contrived in science for urgent telegrams
and electronic messages of alert in order to attract the attention
of independent observers to important astronomical events and
phenomena, to confirm or deny the incoming information. The presence
of eclipses and the precessing accretion disk in the system
HZ~Her, as well as the absence of eclipses in the system V1341~Cyg
are not the "model representations," but objective facts, firmly
established by the observations at different wavelengths at the
ground-based and space-based observatories.

This paper describes the properties of HZ~Her and V1341~Cyg, which
are based on the genuine observations.

\begin{acknowledgements} 

We used the database of the American Association of Variable Star
Observers AAVSO, the Northern Sky Variability Survey NSVS,
the SuperWASP wide-angle search for exoplanets, as well as the
electronic version of the General Catalog of Variable Stars.
The operation of the SAO RAS telescopes are funded by the Ministry
of Education and Science of the Russian Federation.
\end{acknowledgements}

\begin{figure*}[p]
\includegraphics[scale=0.45]{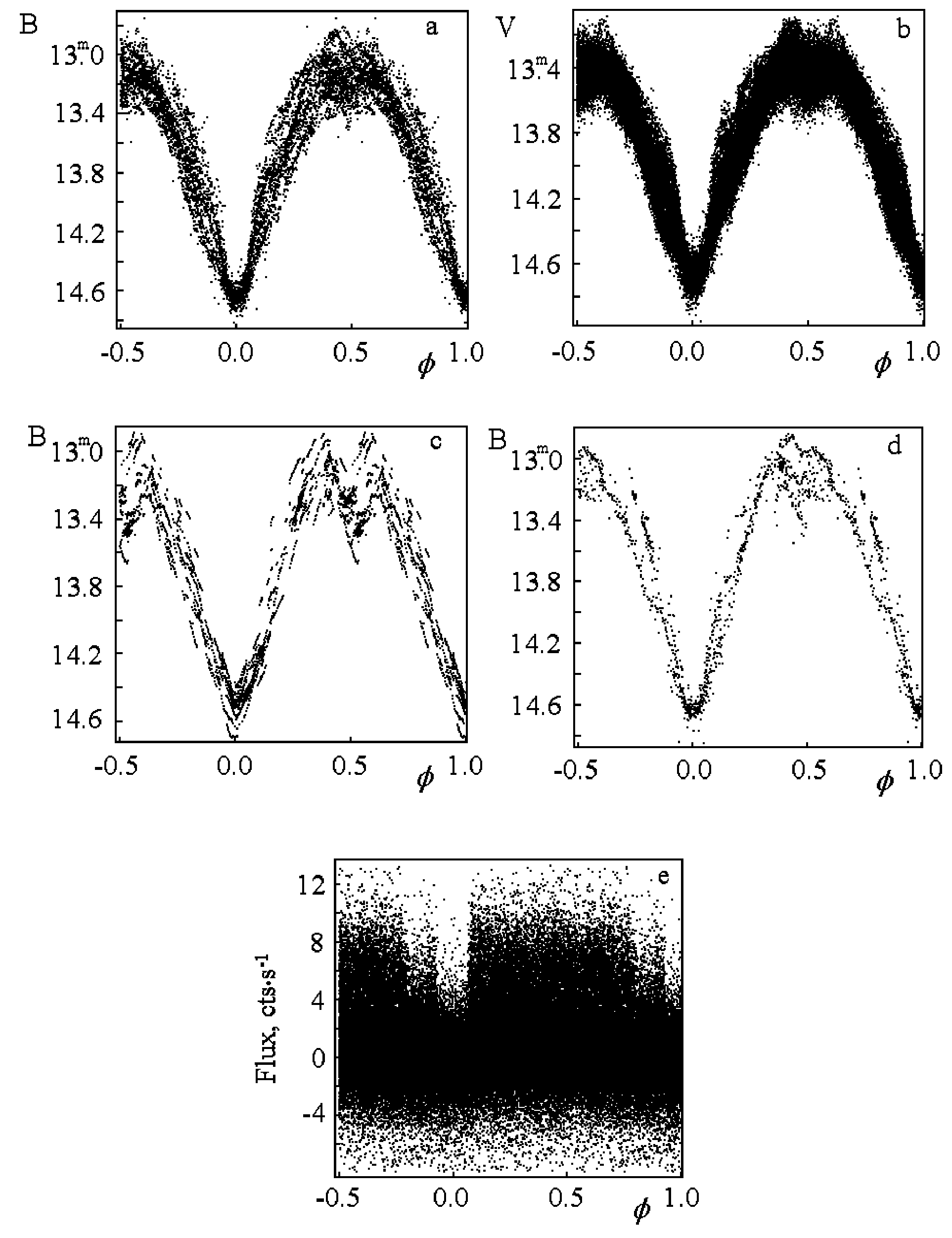}
\caption{
Orbital light curves of HZ~Her in the optical range
(a---\cite{Sha09}, b---\cite{Jur01}) and in the X-ray range (d).
(c) is the light curve according to A. Sazonov~\cite{Saz04}.
(e) is the light curve according to the observations~\cite{Sha09},
built for the sample from the same time range, which includes the data
from~\cite{Saz04}.
}
\end{figure*}

\begin{figure*}[p]
\includegraphics[scale=0.45]{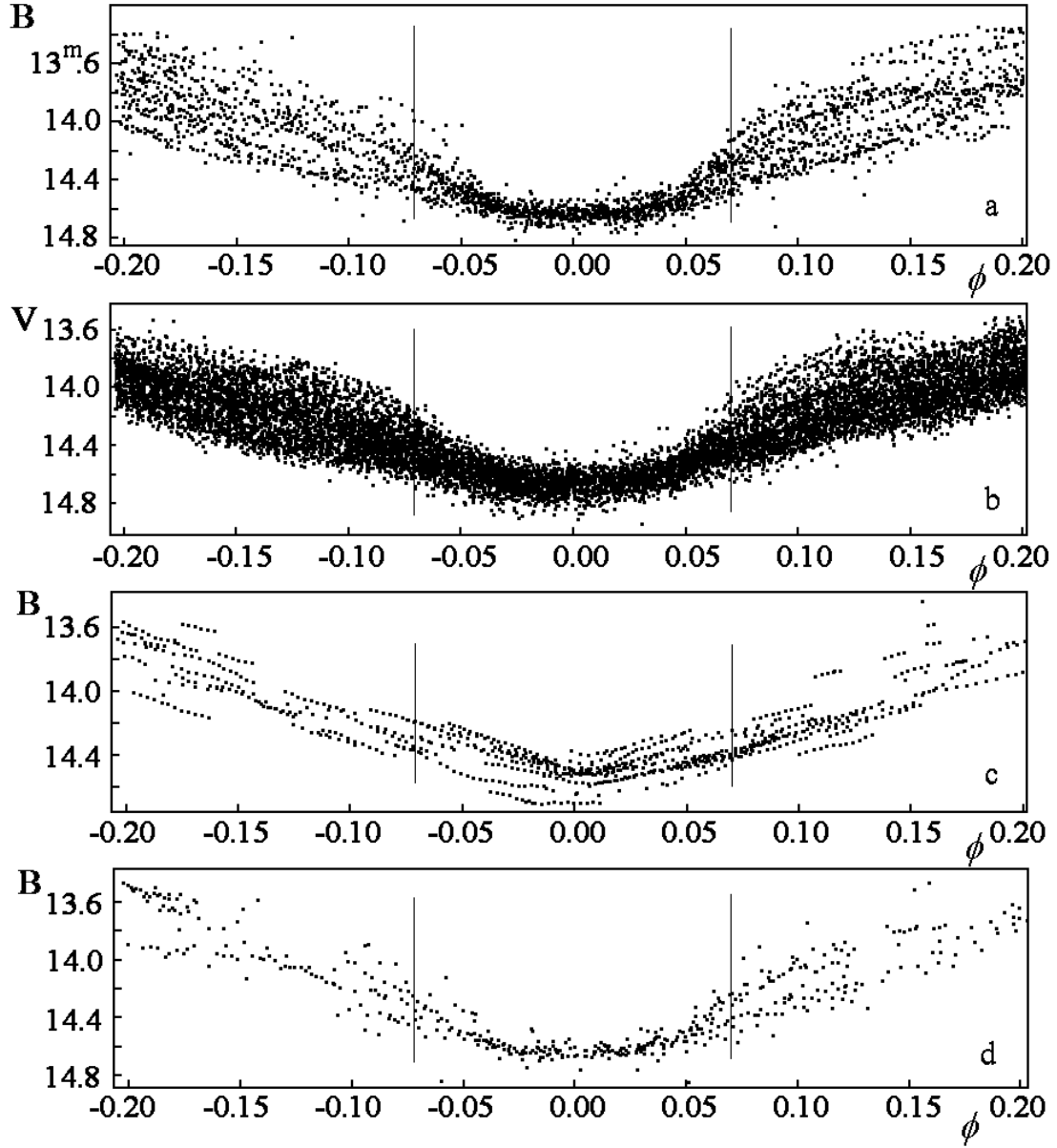}
\caption{
The primary eclipse of HZ~Her in the optical range (a--\cite{Sha09},
b--\cite{Jur01}). (c) the light curve according to A. Sazonov~\cite{Saz04}.
(d) is the light curve  from the observations of~\cite{Sha09}, built for the
sample from the same time range, which includes the data from~\cite{Saz04}.
The vertical lines mark the times of contact of the X-ray pulsar
-- the compact X-ray source -- with the limb of the secondary
component.
}
\end{figure*}

\begin{figure*}[p]
\includegraphics[scale=0.4]{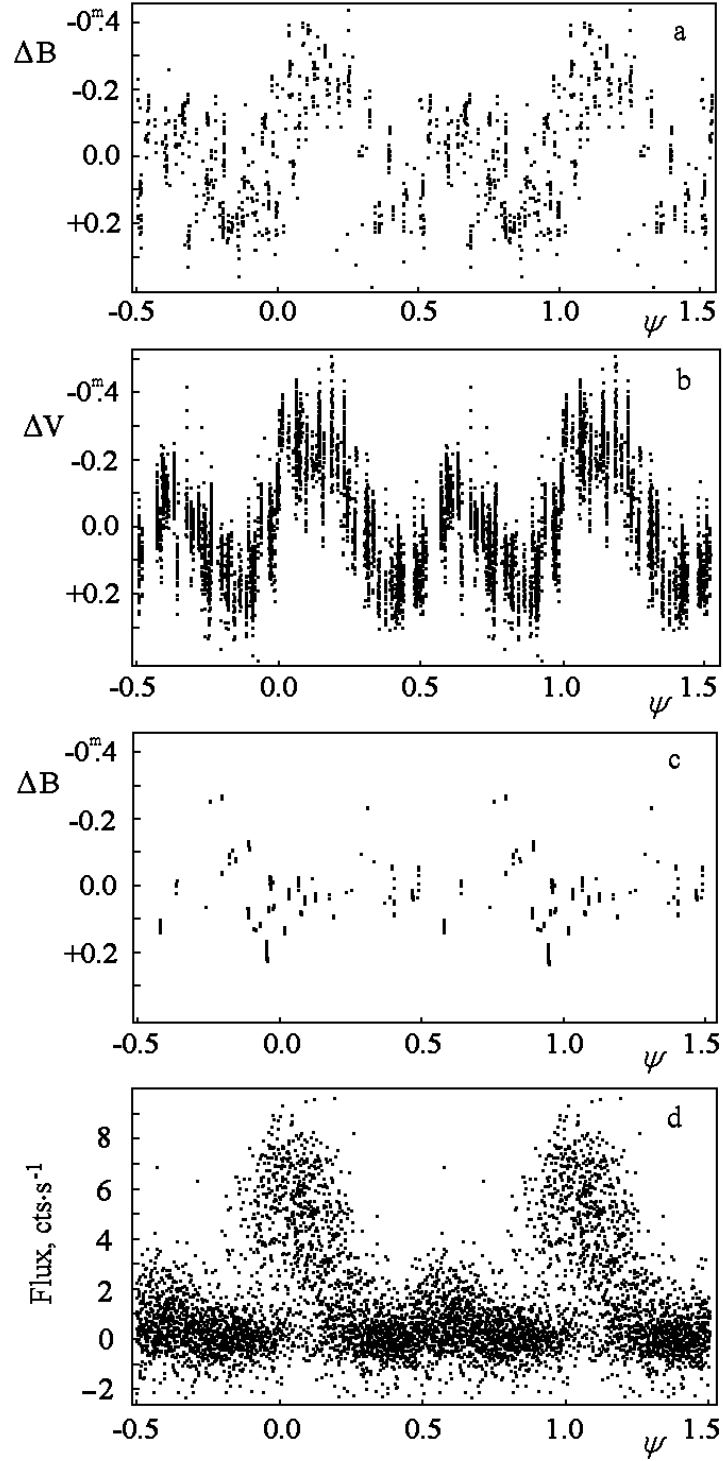}
\caption{
Precession brightness variations of HZ~Her in the orbital
period  phases $\varphi 0.86 - 0.92$ and $0.08 - 0.14$ (a--\cite{Sha09},
b--\cite{Jur01}); (c) is the light curve according to  Sazonov~\cite{Saz04};
(d) -- is the $2 - 10$~keV X-ray flux curve according to the RXTE/ASM
data.
}
\end{figure*}

\begin{figure*}[p]
\includegraphics[scale=0.45]{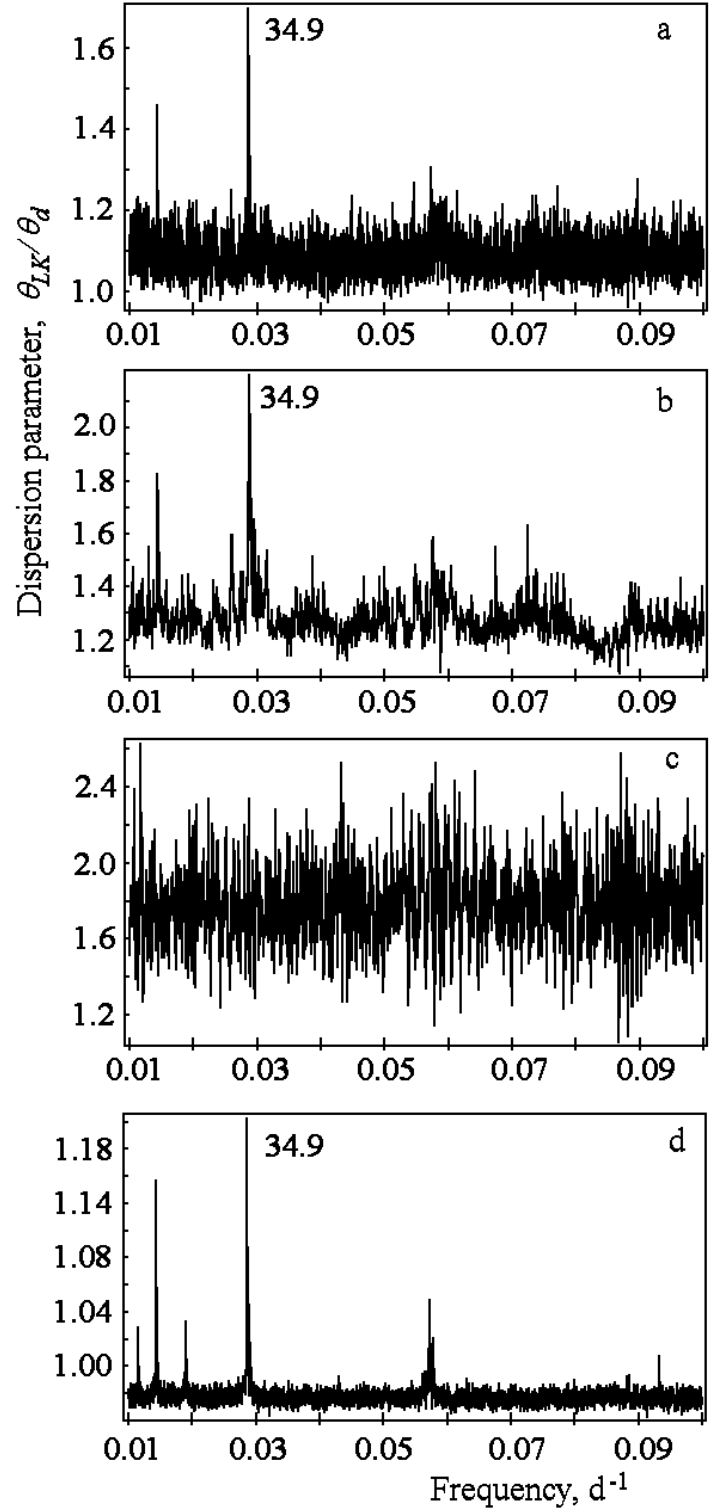}
\caption{
Periodograms of the search of the secondary period in in
the orbital light curve shape variations of HZ~Her by
Goranskij~\cite{Gor21} (a---according to~\cite{Sha09}, b---from~\cite{Jur01},
c---according to the RXTE/ASM data). Sazonov's data (c) lacks periodic changes
in the orbital light curve shape.
}
\end{figure*}

\begin{figure*}[p]
\includegraphics[scale=0.45]{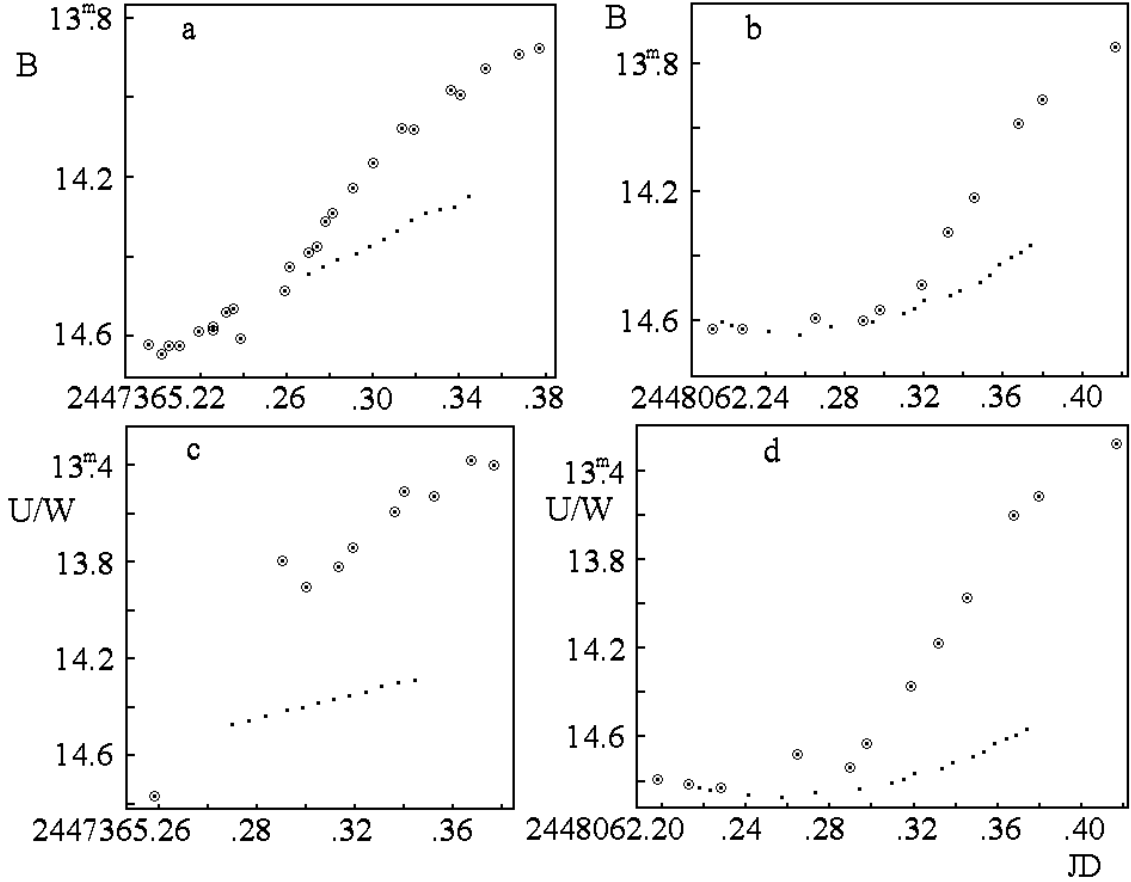}
\caption{
A comparison of the observations of Kilyachkov et al.~
\cite{Kil12} coinciding in time (dots in circles) and the data from the
paper by Sazonov~\cite{Saz04} (dots) in the $B$ filter (top) and in the UV
filters $U$ and $W$ (bottom) over two nights in which the object was
in its "on" state.  Sazonov's observations do not reflect the "on"
state at the end of the eclipse, i.e. do not contain information
on the uncovering of the accretion disc.
}
\end{figure*}

\begin{figure*}[p]
\includegraphics[scale=0.45]{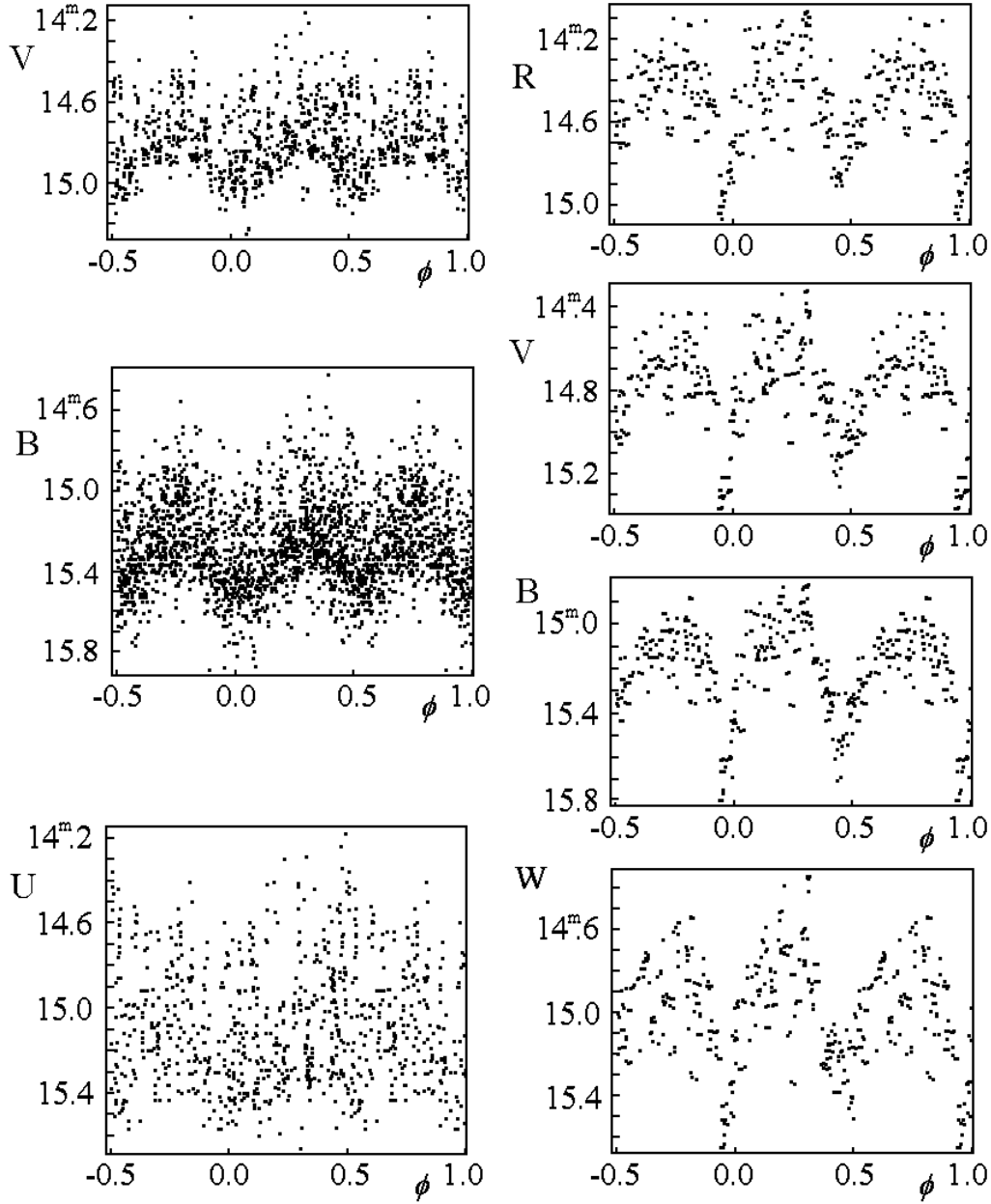}
\caption{
Orbital brightness variations of V1341~Cyg in the $UBV$
system filters according to the data of Goranskij's collection
described above (left) and in the $WBVR$ system filters according to
Sazonov~\cite{Saz05} (right). The scale of all the diagrams is the same.
}
\end{figure*}

\begin{figure*}[p]
\includegraphics[scale=0.5]{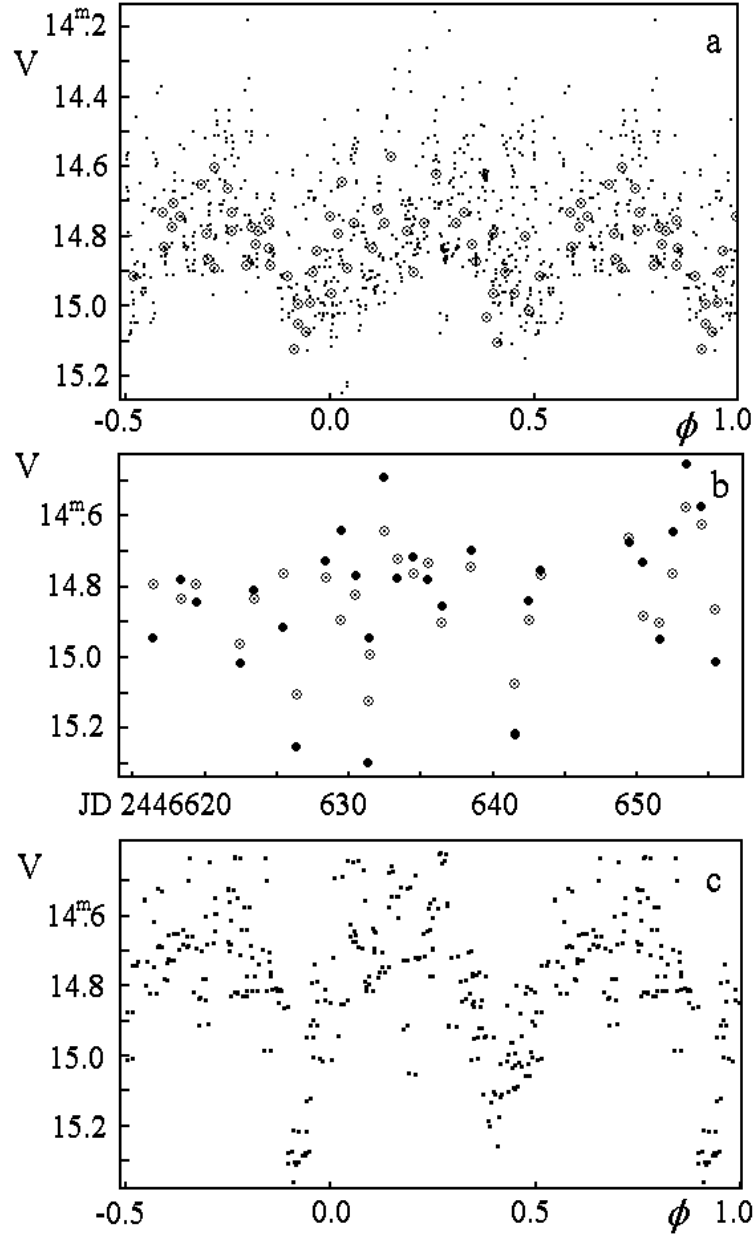}
\caption{
(a) A comparison of observations of V1341~Cyg in the V-band
from the collection by Goranskij (points) with the observations of
the early work of Sazonov~\cite{Saz13} (points in circles). A good
agreement of these data is obvious. (b) A comparison of the early
observations by Sazonov from~\cite{Saz13} (points in circles) with his own
observations from~\cite{Saz05} (circles). Obviously, these are the same
observations, coinciding in time, but later they were amended to
increase the amplitude. (c) The corrected light curve from~\cite{Saz05}
already looks like an eclipsing one.
}
\end{figure*}

\begin{table}[!p]
\caption{Determination of the periods of the orbit, beats and precession
of HZ~Her}
\label{tab1}
\bigskip
\begin{tabular}{|c|c|c|c|c|}
\hline
Parameter & Shakura et al. & Sazonov        & SuperWASP      &  RXTE \\
          &   ($V$)        &    ($B$)       &    ($V$)       &       \\
\hline
JD 24...  & 41511 -- 51046 & 46588 -- 48173 & 53129 -- 54681 & 50087 -- 55817 \\
\hline
$P_{orb}$ & 1.70017 (1) LK & 1.70014 (6) LK & 1.70016 (3) LK & 1.70017 (4) LK\\
\hline
$P_{beat}$& 1.62116 (2) D  &     no         & 1.6210 (2) D   & 1.6213 (1) D \\
\hline
$P_{prec}$& 34.89 (1) G    &     no         & 34.74 (4) G    & 34.95 (2) D, LK \\
          & 34.89 (1) LK*  &     no         & 34.75 (5) LK*  & 34.5 -- 35.2  O--C \\
\hline
$N$       & 5936           & 1808           & 32143          & 95558  \\
\hline
\multicolumn{5}{l}{* \footnotesize
For the phase-selected extraction near ingress and egress of the eclipse.}
\end{tabular}%
\end{table}

\begin{table}[!p]
\caption{Determination of the orbital period of V1341~Cyg}
\label{tab2}
\bigskip
\begin{tabular}{|c|c|c|c|c|}
\hline
Parameter &  Filter      & Goranskij et al. &    Sazonov     & SuperWASP  \\
\hline
JD 24...  &    --        &  42247 -- 55826  & 46616 -- 48521 & 54279 -- 54419 \\
\hline
$P_{orb}$ &   $U$        &    no            & 9.843(2)       & \\
\hline
$P_{orb}$ &   $B$        & 9.8445(4)        & 9.842(2)       & \\
\hline
$P_{orb}$ &   $V$        & 9.8447(4)        & 9.842(2)       &   9.82(7) \\
\hline
\end{tabular}
\end{table}

\end{document}